\documentclass{article}
\usepackage{amsmath}
\usepackage{amssymb}
\usepackage{amsthm}
\usepackage{graphicx}
\usepackage{psfrag}
\usepackage[hang, nooneline]{subfigure}

\author{Natascha Riahi \footnote{e-mail address: riahi@ap.univie.ac.at}
\\ University of Vienna, Faculty of Physics, Gravitational Physics
\\ Boltzmanng. 5, 
1090 Vienna, Austria}
\title{Solving the time-dependent Schr\"odinger equation via Laplace Transform}
\begin{document}
\maketitle
\begin{abstract}
We show how the Laplace transform can be used to give a solution of the time-dependent 
Schr\"odinger equation for an arbitrary initial wave packet
if the solution of the stationary equation is known. The solution is obtained without
summing up eigenstates nor do we need the path integral.
We solve the initial value problem for three characteristic piecewise constant potentials.
The results give an intuitive picture of the wave packet dynamics,
reproducing explicitly all possible reflection and transmission processes. We investigate 
classical and
quantum properties of the evolution and determine the reflection and transmission probabilities.
\end{abstract}

\section{Introduction}
The most elementary and standard way to solve the time-dependent Schr\"odinger equation  is to express the 
initial 
wave function in terms of the eigenfunctions and write the solution as a sum of
and/or an integral over the corresponding eigensolutions oscillating with the corresponding frequencies.
This method is appropriate to investigate the revival behaviour \cite{Ro} which  is a pure quantum
phenomenon that occurs in many quantum systems at late times.
But in order
to get information about details of the wave packet dynamics including the transition from 
classical to quantum mechanics taking place much earlier than the revival phenomenon
the eigenstates have to be summed up numerically \cite{Square},\cite
{Hydro}. Moreover there are cases
such as the  finite square well where the energy eigenvalues  $E(n)$ 
are only given implicitly \cite{FSquare}, which complicates the summation.

An alternative route is the Feynman path integral that contains
the classical action and seems to be a good candidate for
the study of quasiclassical behaviour.
But even attempts to simplify the path integral via semiclassical approximations such 
as the Gutzwiller trace formula
\cite{Gutz} or the 
method of cellular dynamics \cite{Heller} still lead to  involved expressions. Moreover 
evaluating these 
semiclassical formulas
one can never be sure how long they are a good approximation
for the exact behaviour of the wave function.

So the question arises wether there is something in between the brute force method of 
summing up eigenstates 
and entering the world of path integrals. 
For the infinite square well a third possibility is known.
The so-called mirror solution \cite{Kleber} consists of a sum of free particle solutions, describing 
a wave packet (particle)
being reflected from one wall to the other. This solution is achieved by just considering the special symmetry of 
the problem and making a successful guess. So it would be desirable
to have a method to derive such intuitive solutions also for more
general problems.

The method of Laplace transform  that we will introduce
gives the exact solution of the time-dependent Schr\"odinger equation when the solution of the stationary equation is known. 
For our purpose it is not necessary to identify the eigenvalues.
Instead the Laplace transformed Schr\"odinger equation contains the initial wave packet as an 
inhomogeneous term  which allows to apply the method of variation of constants. 
The solution is then given in terms of the initial wave packet.
 The only technical
difficulty is to perform the inverse Laplace transform. Here we can
use the toolbox of rules and inversion formulas, provided for instance in
\cite{Za},\cite{Erdely}.

Though  the technique of Laplace transformation has been applied to
several specific problems concerning the time dependent Schr\"odinger equation (see below), 
the proposed method has 
not yet been consequently used and explored as a tool to solve the initial 
value problem of wave packet evolution.
 
In \cite{Eyring} and \cite{Hladik} the Laplace transform was used to 
determine a perturbative solution of the time 
dependent 
Schr\"odinger equation with the help of an eigenfunction expansion of the unperturbed Hamiltonian. 
In \cite{Arnold} and \cite{Villa} the Laplace transform was applied to problems with time 
dependent boundary conditions.
   \cite{Feld},\cite{Garcia} explored the late time 
behaviour of  special initial wave functions for tunneling phenomena.
Both authors required the  identification of complex poles of the Laplace transformed solution  which 
brings back a 
problem of similar 
difficulty as the identification of eigenvalues 
and makes the solutions less explicit.

What comes closer to the proposed method is the derivation of the propagator $K(x,x',t)$ from the Green 
function which means to invert the  half-sided Fourier transform (\cite{Moretti}, \cite{Krylov})
\[
G(x,x',E)=\int_{0}^{\infty}K(x,x',t)e^{i E t}dt \,.
\]
Since this integral can in general not be expected to converge  this implies a regularization 
$E \rightarrow E+i \epsilon$ .
This makes the Laplace transform favorable, where such a procedure is not needed and
the existing inversion formulas can be applied.

We start by introducing the method in section \ref{Method}. 
In section \ref{ISW} we show that we can reproduce the mirror solution of the infinite square well in a 
straightforward way.
We proceed with the potential step in section \ref{PS}. 
The derivation of the exact solution in this case 
is significantly easier
than the derivation of the propagator in \cite{Carvalho},\cite{ Krylov} where the PDX method for 
path integrals was 
applied. Apart from being of similar simplicity as the most explicite solutions available in the 
literature
(\cite{Los},\cite{ Krylov}) the obtained solutions admit a direct insight into the wave packet dynamics.
We show directly that sufficiently peaked wave packets with energy higher than the 
 potential step will
 be slowed down as a corresponding classical particle which was already derived in \cite{Moretti} using 
 the 
 saddle point
 approximation. And we investigate the wave packet behaviour in the classically forbidden region 
 for wave packets
 with energy lower than the energy of the step. We are able to derive an exact result for the reflection
 probability in terms of the transmission coefficient of the time independent solutions (\ref{Finding}).
 This result confirms a conjecture 
 in \cite{Krylov2}
 under a certain condition for the initial wave packet.

In section \ref{ASW} we investigate the asymmetric square well, a potential
 that is a combination of the two preceding ones and can be
described as a box with exit. As far as we know the 
 initial value problem of this model was not considered before. We  find out that the 
 solution  is a very intuitive combination of 
 the solutions of the infinite square well and the potential step. We show that for
 reasonably peaked wave packets and after not too long times the probability of a particle 
 to be found in the box is given by powers of  the  reflection  probability of the 
 potential
 step, where the power is given by
 the number of
 reflections the wave packet has already undergone.
\section{The method}
\label{Method}
The Laplace transform \cite{Za} $\varphi(x,s)$ of the wavefunction $\psi(x,t)$ reads

\begin{equation}
\varphi(x,s) = \mathcal{L}(\psi(x,t)) = \int\limits_{0}^{\infty} \psi(x,t)  e^{-s t} dt \quad .
\end{equation}

Here and from now on we  always assume that the Laplace transform of the
wave function exists which is ensured if the wavefunction does not grow
faster than exponentially in time. 

Applying the Laplace transformation to the Schr\"odinger equation

\begin{equation}
-\frac{\hbar^2}{2 m} \frac{\partial^2 \psi(x,t)}{\partial x^2} + V(x) \,
\psi(x,t) = i  \hbar \,   \frac{\partial \psi(x,t)}{\partial t}
 \end{equation}

we find

\begin{equation}
\label{Wampi}
-\frac{\hbar^2}{2 m} \frac{\partial^2 \varphi(x,s)}{\partial x^2} + V(x) \,
\varphi(x,s) = i \hbar \, s \,  \varphi(x,s) - i \hbar \, \psi(x,0) \,,
\end{equation}
 where $\psi(x,0)$ is the initial wave function.

Equation (\ref{Wampi}) is an inhomogeneous linear ordinary second order differential
equation for the Laplace-transformed wave function $\varphi(x,s)$. When two
linearly independent solutions of the homogeneous equation are known a
particular solution of (\ref{Wampi}) can be achieved by the method of
variation of constants. With the two homogeneous solutions $u_{1}(x,s)$ and
$u_{2}(x,s)$ we get the particular solution
\begin{equation}
\varphi_{p}(x,s) = u_{1}(x,s) \int\limits_{c}^{x}
\frac{2m i}{\hbar} \,\frac{u_{2}(y,s)}{W[u_{1},u_{2}]} \, \psi(y,0) \,dy \, -
 \,u_{2}(x,s) \int\limits_{c}^{x}
\frac{2m i}{\hbar} \,  \frac{u_{1}(y,s)}{W[u_{1},u_{2}]} \, \psi(y,0) \,dy
\end{equation}
Here $W[u_{1},u_{2}]$ is the Wronskian of the two homogeneous solutions:
\[
W[u_{1},u_{2}] = \frac{\partial u_{1}(x,s)}{\partial x}\,u_{2}(x,s) -
 \frac{\partial u_{2}(x,s)}{\partial x}\,u_{1}(x,s) \quad.
\] 
Therefore the general solution of (\ref{Wampi}) reads
\begin{equation}
\varphi_(x,s) = \varphi_{p}(x,s)+\alpha(s)\, u_{1}(x,s)+\beta(s)\, u_{2}(x,s) \quad
\end{equation} 
The functions $\alpha(s)$ and  $\beta(s)$ are determined by the boundary conditions on $\varphi(x,s)$,
which stem from the physical boundary conditions on the wave function $\psi(x,t)$.  What is left is working out
the inverse Laplace transformation for the specific problem.

\section{The infinite square well}
\label{ISW}
The infinite square well is a physical model system which can be interpreted as the 
one-dimensional description of a particle locked up by infinitely rigid walls.
The appropriate potential is

\[
\begin{array}{lll}

V(x)=\infty  & \mbox{for} & x \leq\,0
\\ V(x)=0 & \mbox{for} & 0\,<\,x\,<\,d
 \\  V(x)=\infty & \mbox{for} & x \geq\,d\quad.
 \end{array}
\]

The dynamics of a wave packet is therefore determined by
\begin{subequations}
\begin{equation}
\label{Kastena}
-\frac{\hbar^2}{2 m} \frac{\partial^2 \psi(x,t)}{\partial x^2}
 = i  \hbar \,   \frac{\partial \psi(x,t)}{\partial t} \,,\quad 
 \mbox{for} \quad x\, \epsilon\, [0,d]
\end{equation}

with the boundary conditions

\begin{equation}
\psi(0,t)=\psi(d,t)=0 \,.
\end{equation} 
\end{subequations}

\begin{subequations}
The Laplace transformed wave packet fulfills 
\begin{equation}
\label{KastenL}
-\frac{\hbar^2}{2 m} \frac{\partial^2 \varphi(x,s)}{\partial x^2} = 
i \hbar \, s \,  \varphi(x,s) - i \hbar \, \psi(x,0) 
\end{equation}

with the boundary conditions 

\begin{equation}
\label{boundaryK}
\varphi(0,s)=\varphi(d,s)=0 \quad.
\end{equation}

\end{subequations}

The  solution of the homogeneous equation corresponding to (\ref{KastenL}) is a linear combination 
of the functions

\[
u_{1}(x,s)=e^{i\sqrt{\frac{2 m s i}{\hbar}} x}\qquad 
u_{2}(x,s)=e^{-i \sqrt{\frac{2 m s i}{\hbar}} x}\quad.
\]

With the Wronskian $W[u_{1},u_{2}]=2i \sqrt{\frac{2 m s i}{\hbar}}$ the general solution 
of (~\ref{KastenL}) reads 

\begin{align}
\label{LoesungKA}
\varphi(x,s)&=\sqrt{\frac{m}{2 i s \hbar}} \left \{u_{1}(x,s)\int_{0}^{x}u_{2}(y,s) \psi(y,0) dy \, - \,
u_{2}(x,s)\int_{0}^{x}u_{1}(y,s) \psi(y,0) dy  \right\} \nonumber \\
\nonumber \\
&\quad +\, \alpha(s) u_{1}(x,s)  \,+\, \beta(s) u_{2}(x,s)  \,\,.  
\end{align}

Solving the equations given by the boundary conditions (\ref{boundaryK}) 
we find for $\alpha(s)$ and $\beta(s)$

\begin{align*}
&\alpha (s)\,=
\sqrt{\frac{m}{2 i s \hbar}} \left 
\{\frac{u_{1}(d,s)}{u_{2}(d,s)-u_{1}(d,s)}
\int_{0}^{d}u_{2}(y,s) \psi(y,0) dy \right. \\
& \qquad \qquad - 
\left.\frac{u_{2}(d,s)}{u_{2}(d,s)-u_{1}(d,s)}
\int_{0}^{d}u_{1}(y,s) \psi(y,0) dy  \right\}=\\
&-\sqrt{\frac{m}{2 i s \hbar}}\int_{0}^{d}u_{1}(y,s) \psi(y,0) dy+\\
&\sqrt{\frac{m}{2 i s \hbar}}\{\frac{u_{1}(d,s)}{u_{2}(d,s)-u_{1}(d,s)}
\left\{\int_{0}^{d}u_{2}(y,s) \psi(y,0) dy 
-\int_{0}^{d}u_{1}(y,s) \psi(y,0) dy \right\}
\\
&\beta (s)\,=-\alpha(s) \quad.
\end{align*}

Inserting the result into (\ref{LoesungKA}) and rewriting the integrals so that each term goes to zero for
$s \rightarrow \infty$ (which is a necessary condition for the successfull application of
the inverse Laplace transform), we get

\begin{eqnarray}
\label{KastenB}
\varphi(x,s)=\frac{\kappa}{\sqrt{2 is }}
\left\{\int_{0}^{x} e^{(i-1)\kappa \sqrt{s} (x-y)} \psi(y,0) dy 
\,+ \,
\int_{x}^{d} e^{(i-1)\kappa \sqrt{s} (y-x)} \psi(y,0)dy \right.\nonumber \\
\left.
\,-\,\int_{0}^{d} e^{(i-1)\kappa \sqrt{s} (x+y)} \psi(y,0) dy \right\}
+
\frac{\kappa}{\sqrt{2 i}}\, s\,  w (s)\, \int_{0}^{d}
\left\{ e^{(i-1)\kappa \sqrt{s} (2d+x+y)} \right. \nonumber \\
\left. \,-\,
 e^{(i-1)\kappa \sqrt{s} (2d+x-y)}\,+\, e^{(i-1)\kappa \sqrt{s} (2d-x-y)} \,-\,
e^{(i-1)\kappa \sqrt{s} (2d-x+y)}\right\}\psi(y,0)\,dy 
\end{eqnarray}

where we have used the abbreviation 

\begin{equation}
\label{Kappa}
\kappa = \sqrt{\frac{m}{\hbar}}
\end{equation}

and $w(s)$ stands for
\[
w(s)=\frac{1}{s^\frac{3}{2}(e^{2 (i-1)\kappa d\sqrt{s}}-1)} \qquad .
\]
Writing $w(s)$ as  a geometric series
\begin{equation}
\label{wSumme}
w(s) = \frac{1}{s^{\frac{3}{2}}(e^{2 (i-1)\kappa d\sqrt{s}}-1)}=
-\frac{1}{s^{\frac{3}{2}}}\sum_{k=0}^{\infty} e^{2 (i-1)d \kappa \sqrt{s}\, k}
\end{equation}
and using the inverse Laplace transform \cite {Erdely}
\begin{equation}
\label{Wurzel1}
\mathcal{L}^{-1}\left\{\frac{1}{\sqrt{s}}e^{-(1-i) \sqrt{a s}}\right\}=
\frac{e^{\frac{i a}{2 t}}}{\sqrt{\pi t}} 
\qquad  \mbox{where} \qquad a\,>\,0 
\end{equation}
   we find for the solution of (\ref{Kastena}) the so-called mirror solution (\cite{Kleber})
\begin{align}
\label{DynKb}
\psi(x,t)= \frac{\kappa}{\sqrt{2 \pi t \,i}} \left( 
\int_{0}^{d} \{ e^{\frac{i (x-y)^2 \kappa ^2}{2 t}} 
-  e^{\frac{i (x+y)^2 \kappa ^2}{2 t}}\} \psi(y,0) dy \right.\nonumber \\
\,+\, \sum_{k=1}^{\infty} \int_{0}^{d} 
\{ e^{\frac{i \kappa^2 (2dk+x-y)^2}{2 t}}- e^{\frac{i \kappa^2 (2dk+x+y)^2}{2 t}}  \nonumber \\
 + \left.\,e^{\frac{i \kappa^2 (2dk-x+y)^2}{2 t}}- e^{\frac{i \kappa^2 (2dk-x-y)^2}{2 t}}\} \psi(y,0) dy \right)\,.
\end{align}

At times much smaller than the revival time ($T_{rev}=4 m d^2/\hbar$) of the infinite square well, a given wave packet is reflected between the walls with the classical period corresponding to its momentum expectation value.
At a certain period of motion, a certain term of (\ref{DynKb}) is dominant. 
The first term is just the free particle propagator. When the wave packet, initially located between the walls,
leaves the box it will no longer yield a relevant contribution to the solution. The second term represents
the free dynamics of a wave packet, that has the initial shape $\psi(-x,0)$, since
\begin{equation}
\label{reflection}
\int_{0}^{d} e^{\frac{i (x+y)^2 \kappa ^2}{2 t}} \psi(y,0) dy =
\int_{-d}^{0} e^{\frac{i (x-y)^2 \kappa ^2}{2 t}} \psi(-y,0) dy \,.
\end{equation}
Furthermore, if $\psi(x,0)$ has the momentum expectation value $p_{0}$, the mirrored wavefunction $\psi(-x,0)$
has just the opposite momentum $-p_{0}$.
Therefore, if the first term describes a wave packet moving towards the left wall, the second describes
a wave packet initially located left outside the box and entering it around the time when the classical 
reflection takes place.
So the first two terms  of (\ref{DynKb}) are the quantum version of the reflection of a classical particle 
at the left wall.
Similarly, each term can be given a certain meaning in terms of classical 
dynamics. If the initial wave packet  moves towards the right wall, the term (\ref{reflection})
becomes negligible, and another term of (\ref{DynKb}) will be responsible for the description
of the first reflection at the right wall.
For each of the subsequent reflections there will be
two relevant terms describing the reflection process.
These terms coincide with the exact solution for a wave packet 
reflected at a single infinite wall which was already investigated in detail 
(\cite{Bounce1},\cite{Bounce2},\cite{Bounce3}).
As time progresses, of course the spreading of wave packets increases and the semiclassical picture is no 
longer
correct.

\section{The potential step}
\label{PS}
Out next model system is the potential step, described by the potential

\[
\begin{array}{lll}
V(x)=0 & \quad \mbox{for} \quad & x\,<\,0
\\  V(x)=V & \quad \mbox{for} \quad & x\,\geq\,0 \quad.
 \end{array}
\]

The evolution of the wave packet $ \psi(x,0) $, is determined by

\begin{align}
\label{Stufea}
&-\frac{\hbar^2}{2 m} \frac{\partial^2 \psi(x,t)}{\partial x^2}
 = i  \hbar \,   \frac{\partial \psi(x,t)}{\partial t} \qquad &
 \mbox{for} \qquad  x\,<\,0 \\
& -\frac{\hbar^2}{2 m} \frac{\partial^2 \psi(x,t)}{\partial x^2}+V \psi(x,t)
 = i  \hbar \,   \frac{\partial \psi(x,t)}{\partial t} \qquad &
 \mbox{for} \qquad  x\,\geq\,0\,,
 \end{align}
 
 where $\psi(x,t)$ is supposed to be continuously differentiable. 
 The solution will be square integrable for all times, if the initial function is chosen square integrable. We will from now on only consider initial wave packets
 that are located left of the potential step
 
 \[
  \psi(x,0)=0 \qquad \mbox{for}\qquad x \geq 0\,.
 \]
 
 According to our method, we find for the Laplace transformed wave packet $\varphi(x,s)$

\begin{subequations}
\label{LoesungStufe}
\begin{align}
\varphi(x,s)&=\sqrt{\frac{m}{2 i s \hbar}} \left \{u_{1}(x,s)\int_{-\infty}^{x}u_{2}(y,s) 
\psi(y,0) dy \,  - \,\right.&
\left.
 u_{2}(x,s) 
\int_{-\infty}^{x}u_{1}(y,s) \psi(y,0) dy 
\right\} \nonumber \\
& &\nonumber \\
&\quad +\, \alpha(s) u_{1}(x,s)  \,+\, \beta(s) u_{2}(x,s) 
 &\mbox{for}  \quad x\,<\,0 \qquad\\
& &\nonumber \\
\varphi(x,s)&= \gamma(s) u_{3}(x,s)  \,+\, \delta(s) u_{4}(x,s)
 &\mbox{for}  \quad x\,>\,0 \qquad ,
\end{align}
\end{subequations}

where
\begin{subequations}
\label{u1234}
\begin{align}
& u_{1}(x,s)=e^{i\sqrt{\frac{2 m s i}{\hbar}} x}\qquad 
& u_{3}(x,s)=e^{i\sqrt{\frac{2 m s i}{\hbar}-\frac{2 m V}{\hbar^2}} x} 
\\
& u_{2}(x,s)=e^{-i \sqrt{\frac{2 m s i}{\hbar}} x}\qquad
& u_{4}(x,s)=e^{-i \sqrt{\frac{2 m s i}{\hbar}-\frac{2 m V}{\hbar^2}} x}\,.
\end{align}
\end{subequations}

Since $\psi(x,t)$ must be square integrable, $\varphi(x,s)$ must vanish for $x \rightarrow \pm \infty$.
We get
\[
\alpha(s)=\delta(s)=0 \,.
\]

Furthermore $\varphi(x,s)$ should be continuously differentiable at $x=0$ which determines
the functions $\beta(s)$ and $\gamma(s)$ :

\begin{subequations}
\begin{align}
&\gamma(s)=
\sqrt{\frac{m}{2 s i \hbar}}
\int_{-\infty}^{0}u_{2}(y,s) 
\psi(y,0) dy \,
\cdotp\frac{2}{1+\sqrt{1-\frac{V}{\hbar s i}}} \\
&\beta(s)=\sqrt{\frac{m}{2 s i \hbar}}
\int_{-\infty}^{0}u_{2}(y,s) 
\psi(y,0) dy 
\,\cdotp\left(
\frac{2}{ 1+\sqrt{1-\frac{V}{\hbar s i}} }-1
\right)
 \nonumber\\ &\qquad+
\sqrt{\frac{m}{2 s i \hbar}}
\int_{-\infty}^{0}u_{1}(y,s) 
\psi(y,0) dy \,.
\end{align}
\end{subequations}

\begin{subequations}
\label{LStufe}
This yields for the Laplace transformed wave packet $\varphi(x,s)$ (\ref{LoesungStufe})
\begin{align}
& \varphi(x,s)=
 \sqrt{\frac{m}{2 s i \hbar}}\left\{
\int_{-\infty}^{0}
e^{i\sqrt{\frac{2 m s i}{\hbar}} |x-y|} \psi(y,0) dy+ 
\int_{-\infty}^{0}
e^{-i\sqrt{\frac{2 m s i}{\hbar}} (x+y)}\psi(y,0) dy
\cdotp \rho(s)\right\} \nonumber\\
& \qquad \mbox{for}  \quad x\,<\,0 \label{LStufea}\\
 & \varphi(x,s)=
 \sqrt{\frac{m}{2 s i \hbar}}
\int_{-\infty}^{0}
e^{i\sqrt{\frac{2 m s i}{\hbar}-\frac{2 m V}{\hbar^2}} x} 
e^{-i\sqrt{\frac{2 m s i}{\hbar}} y}\psi(y,0) dy 
\,\cdotp (\rho(s)+1)  \nonumber\\
& \qquad \mbox{for}  \quad x\,>\,0 \,, \label{LStufeb}
\end{align}
\end{subequations}
where 
\begin{equation}
\label{LKoeffizient}
 \rho(s)=\frac{2}{ 1+\sqrt{1-\frac{V}{\hbar s i}} }-1\,.
\end{equation}

\subsection{Solution for $x<0$}

The inverse Laplace transform of (\ref{LKoeffizient}) is \cite {Erdely}

\begin{equation}
\label{Koeffizient}
\mathcal{L}^{-1}\left\{\rho(s)\right\}=r(t)=
\frac{1}{t i} J_{1}\left[\frac{V t}{2  \hbar}\right]e^{\frac{-i V t}{2\hbar}}\,,
\end{equation}
where $J_{1}(x)$ denotes the Bessel function.
Since the inverse Laplace transform of a product is given by a convolution \cite{Za}

\begin{align}
\label{Faltung}
& \mathcal{L}^{-1}\left\{ f(s) \cdotp g(s)\right\}=\int_{0}^{t}
 F\left(t-\tau\right)G(\tau)d\tau\,, \\
& \mbox{where} \quad
  \mathcal{L}^{-1}\left\{ f(s)\right\}=F(t)\,,\,
 \mathcal{L}^{-1}\left\{ g(s)\right\}=G(t)\,,\nonumber
 \end{align}
 
 we find for the inverse Laplace transform of the wave packet (\ref{LStufea}) in the region $x\,<\,0$
 
 \begin{align}
 \label{StufekN}
  \psi(x,t)=
  \frac{\kappa}{\sqrt{2 \pi t \,i}} 
\int_{-\infty}^{0}  e^{\frac{i (x-y)^2 \kappa ^2}{2 t}}\psi(y,0) dy
+
\int_{0}^{t}\frac{\kappa}{\sqrt{2 \pi (t-\tau) \,i}} 
\int_{-\infty}^{0}e^{\frac{i (x+y)^2 \kappa ^2}{2 (t-\tau)}} \psi(y,0) r(\tau)dy\, d\tau \,,
 \end{align}

where we have applied (\ref{Wurzel1},\ref{Kappa}). 
The first term describes the free time evolution of the wave packet
$\psi(x,0)$. The second term represents the  reflected wave packet and has the form of a 
wavepacket reflected at an infinite wall (see also the description
at the end of the last section), deformed by a convolution.

If we use the momentum representation 

\begin{equation}
\label{MREp}
 f(p)=\frac{1}{\sqrt{2 \pi \hbar}}\int_{-\infty}^{\infty} \psi(x,0) e^{\frac{-i p x}{\hbar}} dx\,,
\end{equation}

we can rewrite the convolution integral in (\ref{StufekN}) as 

\begin{align}
 &\int_{0}^{t}\frac{\kappa}{\sqrt{2 \pi (t-\tau) \,i}} 
\int_{-\infty}^{0} e^{\frac{i (x+y)^2 \kappa ^2}{2 (t-\tau)}} \psi(y,0) dy\,r(\tau)d\tau = \\
&\int_{0}^{t}\frac{\kappa}{\sqrt{2 \pi (t-\tau) \,i}} 
\int_{-\infty}^{\infty} e^{\frac{i (x+y)^2 \kappa ^2}{2 (t-\tau)}} \psi(y,0) dy\,r(\tau)d\tau = \\
&\int_{0}^{t}\frac{1}{\sqrt{2 \pi \hbar}} \int_{-\infty}^{\infty} e^{\frac{-i p x}{\hbar}}
e^{-\frac{i p^2(t-\tau)}{2m \hbar}}f(p) dp\,r(\tau)d\tau= \\
&\frac{1}{\sqrt{2 \pi \hbar}} \int_{-\infty}^{\infty} e^{\frac{-i p x}{\hbar}}
e^{-\frac{i p^2 t}{2m \hbar}}f(p) R (p, t)dp\,,
\end{align}

where 

\[
 R(p,t)=
 \int_{0}^{t} e^{\frac{i p^2 \tau}{2 m \hbar}-\frac{i V \tau}{2 \hbar}}
 J_{1}\left[\frac{V \tau}{2  \hbar}\right] \frac{d \tau}{\tau i} \,.
\]

We will approximate this function by

\begin{equation}
 \label{Approx}
 R(p,t)\,\approx R(p)=\int_{0}^{\infty} e^{\frac{i p^2 \tau}{2 m \hbar}-\frac{i V \tau}{2 \hbar}}
 J_{1}\left[\frac{V \tau}{2  \hbar}\right] \frac{d \tau}{\tau i}
\,.
\end{equation}

Since the Bessel function fulfills

\[
 \left|J_{1}\left[\frac{V t}{2  \hbar}\right]\frac{1}{t}\right| \leq
 \frac{2  \hbar}{V }\frac{1}{t^{3/2}}\,
 \]
 
 the additional part of the integral
 \[
 \left| \int_{t}^{\infty} e^{\frac{i p^2 \tau}{2 m \hbar}-\frac{i V \tau}{2 \hbar}}
 J_{1}\left[\frac{V t}{2  \hbar}\right] \frac{d \tau}{\tau i} \right|
 \leq \int_{t}^{\infty} \sqrt{\frac{2  \hbar}{V }}\frac{1}{\tau^{3/2}} d\tau=
 2\sqrt{\frac{2  \hbar}{V t}}
 \]

does hardly contribute, if 
\[
 t\gg \frac{ \hbar}{V }\,,
\]

and can therefore be neglected after a very short time. Moreover, if the initial function is reasonably 
peaked the second term of (\ref{StufekN}) does not become  relevant until the wave packet approaches 
the barrier.
So the approximation can always be used when

\[
 t_{R}\gg \frac{ \hbar}{V } \,,
\]

where $t_{R}$ is the time, a classical particle corresponding to the wave packet would need
to
reach the barrier.

The solution for the region $x\,<\,0$ , then reads
 \begin{align}
 \label{StufekNM}
 & \psi(x,t) \approx
  \frac{\kappa}{\sqrt{2 \pi t \,i}} 
\int_{-\infty}^{0}  e^{\frac{i (x-y)^2 \kappa ^2}{2 t}}\psi(y,0) dy\, + \\
& \quad \frac{1}{\sqrt{2 \pi \hbar}} \int_{-\infty}^{\infty} e^{\frac{-i p x}{\hbar}}
e^{-\frac{i p^2 t}{2m \hbar}}f(p) R(p)dp\, \nonumber.
\end{align}

Since the integral (\ref{Approx}) is itself a Laplace integral, we can apply (\ref{Koeffizient})
and find for  $R(p)$

\begin{equation}
 \label{Rk}
 R(p)=\rho\left[-\frac{i p^2}{2 m \hbar} \right]=-1 + 2 k-2 \sqrt{k(k-1)}
 \quad \mbox{with} \quad k=\frac{p^2}{2 m V}\,.
\end{equation}

This function coincides with the reflection coefficient of the time independent solution 
(see for instance \cite{Messiah}).

Defining the reflection probability as
\[
\mathcal{R}(t)=\int_{-\infty}^{0}\left|\Psi(x,t)\right|^2 dx\,,
\]
we are now able to show that
\begin{equation}
\label{Finding}
\lim_{t \rightarrow \infty} \mathcal{R}(t) \leq \int_{-\infty}^{\infty}\left|f(p) R(p)\right|^2 dp\,,
\end{equation}
if the momentum expectation value $p_{0}$ of the initial wavepacket is positive.

The first term of (\ref{StufekNM}) describes the free time evolution of the wavepacket
$\Psi_{F}(x,t)$. For positive momentum expectation values we find
\begin{equation}
\label{Argument}
 \lim_{t \rightarrow \infty}\int_{-\infty}^{0}\left|\Psi_{F}(x,t)\right|^2 dx=0\,.
\end{equation}
This result can be derived by determining the correction that has to be applied to a free wave packet 
solution
with  $p_{0}=0$ if the expectation value changes to  $p_{0}>0$.
So only the second term  of (\ref{StufekNM}) which describes the reflected wave packet 
 contributes 
\[
\lim_{t \rightarrow \infty} \mathcal{R}(t)=
\lim_{t \rightarrow infty} \int_{-\infty}^{0}\left|\Psi_{R}(x,t)\right|^2 dx\,,
\]
where 
\[
 \Psi_{R}(x,t)= \frac{1}{\sqrt{2 \pi \hbar}} \int_{-\infty}^{\infty} e^{\frac{-i p x}{\hbar}}
e^{-\frac{i p^2 t}{2m \hbar}}f(p) R(p)dp=
 \frac{1}{\sqrt{2 \pi \hbar}} \int_{-\infty}^{\infty} e^{\frac{i p x}{\hbar}}
e^{-\frac{i p^2 t}{2m \hbar}}f(-p) R(p)dp\,.
\]

This is the free time evolution of an initial wave packet $R(p) f(-p) $ which yields
\[
\lim_{t \rightarrow \infty} \mathcal{R}(t) \leq 
\int_{-\infty}^{\infty}\left|\Psi_{R}(x,t) \right|^2 dx
= \int_{-\infty}^{\infty}\left|f(p) R(p)\right|^2 dp\,.
\]
\begin{subequations}

If in particular the distortion by $R(p)$ is small enough so that
\begin{equation}
\label{condition}
\int_{-\infty}^{\infty}p \left |f(-p) R(p)\right|^2 dp < 0\,,
\end{equation}
we find, arguing as for (\ref{Argument}),
\begin{equation}
\label{Con}
\lim_{t->\infty}\int_{-\infty}^{0}\left|\Psi_{R}(x,t)\right|^2 dx=
\int_{-\infty}^{\infty}\left|\Psi_{R}(x,t)\right|^2 dx\,,
\end{equation}

\end{subequations}

which means that the inequality in (\ref{Finding}) is saturated. So the expression for
the  reflection probability  that was conjectured in \cite{Krylov2} turns out to apply only under the 
condition
(\ref{condition}) for the initial wave function. Note that (\ref{Finding},\ref{Con}) are exact
results since (\ref{StufekNM}) tends to the exact solution (\ref{StufekN})for $t \rightarrow \infty$ \,.

\subsection{Solution for $x\,>\,0$}
\label{GN}

In order to determine the inverse Laplace transform of (\ref{LStufeb}) we use the momentum
representation of $\psi(x,0)$. This yields for the integral

\begin{align}
&  \sqrt{\frac{m}{2 s i \hbar}}
\int_{-\infty}^{0}
e^{i\sqrt{\frac{2 m s i}{\hbar}-\frac{2 m V}{\hbar^2}} x} 
e^{-i\sqrt{\frac{2 m s i}{\hbar}} y}\psi(y,0) dy = \\
&\frac{1}{\sqrt{2 \pi h}}\int_{-\infty}^{\infty}  
e^{i\sqrt{\frac{2 m s i}{\hbar}-\frac{2 m V}{\hbar^2}} x}
\sqrt{\frac{m}{2 s i \hbar}}
\int_{-\infty}^{\infty} 
e^{\frac{i p y}{\hbar}}
e^{i\sqrt{\frac{2 m s i}{\hbar}} |y|}dy f(p)dp = \\
& \frac{1}{\sqrt{2  \pi  \hbar}}
\int_{-\infty}^{\infty}
e^{i\sqrt{\frac{2 m s i}{\hbar}-\frac{2 m V}{\hbar^2}} x} 
\frac{1}{s +\frac{i p^2}{2 m \hbar}}\,f(p) dp \,.
\end{align}

According to \cite{Erdely}

\begin{align}
 \label{Wurzel2}
 &\mathcal{L} ^{-1}\left\{2 (s+b+c)^{-1} e^{-\sqrt{a (s+c)}} \right\} =  \nonumber\\
 & \qquad e^{-c t-b t}
 \left\{
 e^{-i \sqrt{a b}} \mbox{Erfc}
 \left[\sqrt{\frac{a}{4 t}} -i \sqrt{b t}\right]
+ e^{i \sqrt{a b}} \mbox{Erfc}
 \left[\sqrt{\frac{a}{4 t}} +i \sqrt{b t}\right] 
 \right\} \nonumber \\
 & \qquad  \mbox{where} \quad \mbox{Re}[a]\,\geq \,0\,,
\end{align}
and $\mbox{Erfc}(x)$ is the entire function
\[
\mbox{Erfc}(x)=\frac{2}{\sqrt{ \pi}}
\int_{x}^{\infty}e^{-u^2}du\,.
\] 

We find with
\[
a=-\frac{2 m i}{\hbar}x^2\,,
\quad b=\frac{i}{2 m \hbar}\left(p^2-2 m V\right)\,,
\quad c=\frac{i V}{\hbar}
\]
for the inverse Laplace transform of the wave packet (\ref{LStufeb}) 
in the region $x\,>\,0$

\begin{subequations}
\label{StufegN}
\begin{align}
\label{StufegNa}
&\psi(x,t)=\int_{-\infty}^{\infty} K\left(x, p, t\right)f(p)\,dp+
 \int_{0}^{t} \int_{-\infty}^{\infty} K\left(x, p ,t-\tau\right)f(p)\,dp\,r(\tau)\,d\tau\,,
 \\
 & \mbox{with} \nonumber \\
 \label{StufegNb}
 & K(x,p,t)= \frac{1}{2 \sqrt{2 \pi \hbar}} e^{\frac{- i V t}{\hbar}-\frac{i t Z^2}{2 \hbar m}}
 \quad \cdotp
 \nonumber \\
 & \quad 
 \left\{
 e^{-\frac{i x Z}{\hbar}} \mbox{Erfc}
 \left[-i \sqrt{\frac{2 m i}{\hbar\, t}} \frac{x}{2}-i \sqrt{\frac{i\, t}{2 \hbar m}} Z\right]
 +e^{\frac{i x Z}{\hbar}} \mbox{Erfc}
 \left[-i \sqrt{\frac{2 m i}{\hbar\, t}} \frac{x}{2}+i \sqrt{\frac{i\, t}{2 \hbar m}} Z\right]
\right\} \,,
 \nonumber \\
\end{align}
\end{subequations}
where we have again used the convolution theorem (\ref{Faltung}) and we have introduced the abbreviation

\[
 Z=\sqrt{p^2-2 m V} \,.
\]

A further interpretation of (\ref{StufegN}) is only possible
 if we distinguish between momentum distributions that are concentrated around $p_{0}\,>\,\sqrt{2 m V}$, 
 where
 a classical transmission of the barrier would be possible and those which
 are concentrated around $p_{0}\,<\,\sqrt{2 m V}$,
 where the wave packet enters a classically forbidden region.

\subsection{A wave packet climbing the potential step}

We can write any initial wave packet $\psi(x,0)$
in momentum space (\ref{MREp})as

\begin{subequations}
\label{Darstellung}
\begin{equation}
 f(p)=e^{\frac{-i p x_{0}}{\hbar}} F\left(p-p_{0}\right) \quad,
\end{equation}

where $F(p)$ fulfills
\begin{equation}
 \int_{-\infty}^{\infty} F^{*}(p)\,p\, F(p) dp=0 \,,\qquad
  \int_{-\infty}^{\infty} \, F^{*}(p)F'(p) dp=0\,.
\end{equation}
\end{subequations}
The expectation values of $\psi(x,0)$ are then given by
\[
\left\langle\hat{x}\right\rangle=x_{0}\,,
\qquad
\left\langle\hat{p}\right\rangle=p_{0}\,,
\]
and the uncertainties will be denoted by $\Delta x_{0}$ and $\Delta p_{0}$.

We will now consider an initial function  $f(p)$ (\ref{MREp})
almost exclusively
concentrated in a region $p\,>\,p_{m}\,>\,\sqrt{2 m V}$, so that we can assume

\begin{equation}
\label{Rand}
 f(p) \approx 0 \quad 
 \mbox{for} 
 \quad p\,<\,p_{m}, 
\end{equation}

Moreover we require the wave packet to be sufficiently peaked around the momentum
expectation value $p_{0}$, so that

\begin{subequations}
\label{Approxp}

\begin{align}
\label{Approxp1}
& \frac{p_{0}}{ m V}F(p-p_{0}) (p-p_{0})
 \approx 0 \,, \\
\label{Approxp2}
&e^{-\frac{i \sqrt{q^2+2 m V}x_{0}}{\hbar}}
F\left(\sqrt{q^2+2 m V}-\sqrt{q_{0}^2+2 m V}\right) \approx \\
& \nonumber \quad 
F\left(\lambda(q-q_{0})\right)
e^{-\frac{i q_{0 }x_{0}}{\lambda \hbar}
-\frac{i x_{0}\lambda}{ \hbar}(q-q_{0 })
-\frac{i x_{0}\lambda}{2 \hbar q_{0}k_{0}}(q-q_{0 })^2}\,,
\end{align}
where
\[
p_{0}=\sqrt{q_{0}^2+2 m V}\,,\quad
\lambda=\sqrt{1-\frac{2 m V}{p_{0}^2}}
\]
\end{subequations}

Finally the difference $p_{m}-\sqrt{2 m V}$ should be big enough to ensure 
\begin{equation}
\label{Faktor}
 \frac{1}{\sqrt{\frac{p^2}{2 m V}-1}} = O(1)\quad \mbox{for} \quad p\,\geq\,p_{m}
\end{equation}
which means that this factor does not have relevant influence on the approximations.

If we now insert the wave packet into the solution for $x<0$ (\ref{StufekNM}) 
we will use $R(p)\approx R(p_{0})$
since according to (\ref{Approxp1}) and the mean value theorem

\begin{align*}
f(p) R(p)=f(p)R(p_{0})+ f(p)R'(p_{1})(p-p_{0})
\approx f(p)R(p_{0})\,,\\
& \quad \mbox{with}\quad p_{1}\,\epsilon\,(p_{0},p)
\end{align*}

where (\ref{Faktor}) ensures that the derivative of $R(p)$  (\ref{Rk}) will not spoil the approximation.
We find for $x<0$

 \begin{align}
 \label{C1}
 & \psi(x,t)\approx
  \frac{\kappa}{\sqrt{2 \pi t \,i}} 
\int_{-\infty}^{0}  e^{\frac{i (x-y)^2 \kappa ^2}{2 t}}\psi(y,0) dy\, + 
 \frac{R(p_{0})}{\sqrt{2 \pi \hbar}} \int_{-\infty}^{\infty} e^{\frac{-i p x}{\hbar}}
e^{-\frac{i p^2 t}{2m \hbar}}f(p) dp\,= \nonumber\\
\qquad& \frac{\kappa}{\sqrt{2 \pi t \,i}} 
\int_{-\infty}^{0}  e^{\frac{i (x-y)^2 \kappa ^2}{2 t}}\psi(y,0) dy\, + 
R(p_{0})
\frac{\kappa}{\sqrt{2 \pi t \,i}} 
\int_{-\infty}^{0}  e^{\frac{i (x+y)^2 \kappa ^2}{2 t}}\psi(y,0) dy\,.
\end{align}

The solution consists of an incoming and a reflected wave packet.
Within our approximation the reflected wave packet
has the same shape as the wave packet reflected by an infinite barrier
(compare \ref{reflection}), but its probability density is reduced by
a factor $R(p_{0})^2$. This factor only depends on
$k_{0}=p^2_{0}/(2 m V)$ and goes from $1$ to zero when $k_{0}$ goes from $1$ to infinity (\ref{Rk}). If  $p_{0}^2/2m$,  which corresponds to the energy expectation value within our approximation, 
is much higher than the potential step, there is hardly any  reflection at the step.

For the interpretation of the solution for $x>0$ (\ref{StufegN})
we rewrite the error function according to \cite{Magnus}
\begin{equation}
\label{Error}
\mbox{Erfc}\left[\frac{u}{2 z}\right]=
\frac{2 z}{\sqrt{\pi}}e^{-\frac{u^2}{4 z^2}}
\int_{0}^{\infty}e^{-z^2 y^2 -u y } dy\,,
\end{equation}
where we choose 
\[
z=\sqrt{\frac{-i m}{2 \hbar t}} \,,\quad
u=-\frac{i m}{\hbar t}x \mp \frac{i}{\hbar}Z\,.
\]
 This yields

\begin{align*}
  K(x,p,t)=e^{-\frac{i V t}{\hbar}}
 \left\{
 \frac{\kappa}{ \sqrt{2 \pi t \,i}} \frac{1}{\sqrt{2 \pi \hbar }}
\int_{0}^{\infty}  e^{\frac{i (x+y)^2 \kappa ^2}{2 t}}
e^{\frac{i Z y}{\hbar}}
dy\, + \,
\frac{\kappa}{ \sqrt{2 \pi t \,i}} \frac{1}{\sqrt{2 \pi \hbar }}
\int_{0}^{\infty}  e^{\frac{i (x+y)^2 \kappa ^2}{2 t}}
e^{\frac{-i Z y}{\hbar}}dy
\right\}\, .
\end{align*}

Applying (\ref{Approxp1},\ref{Approxp2},\ref{Faktor}) we find (see Appendix \ref{Deformed})
\begin{subequations}
\begin{align}
&\tilde{\psi}(y,0):=
\frac{1}{\sqrt{2 \pi \hbar}}
\int_{\sqrt{2 m V}}^{\infty}
e^{\frac{i Z y}{\hbar}}f(p)dp=
\frac{1}{\sqrt{2 \pi \hbar}}
\int_{\sqrt{2 m V}}^{\infty}
e^{\frac{i \sqrt{p^2-2 m V} y}{\hbar}}f(p)dp
\approx \\
& \approx \,
\frac{1}{\sqrt{2 \pi \hbar}}\int_{-\infty}^{\infty}
e^{\frac{i q y}{\hbar}}
\lambda e^{-\frac{i q_{0 }x_{0}}{\lambda \hbar}
-\frac{i x_{0}\lambda}{ \hbar}(q-q_{0 })
-\frac{i x_{0}\lambda}{2 \hbar q_{0}k_{0}}(q-q_{0 })^2}
F\left(\lambda(q-q_{0})\right)
dq\,.
\end{align}
\end{subequations}
This function is a deformed version of the original wave packet $\psi(x,0)$, characterized by the following  properties:
\begin{align*}
&\left\langle\tilde{\psi}|\tilde{\psi}\right\rangle=\lambda\,,\quad
\frac{1}{\lambda}\left\langle\tilde{\psi}|\hat{p}|\tilde{\psi}\right\rangle=
q_{0}\,,\quad
\frac{1}{\lambda}\left\langle\tilde{\psi}|\hat{x}|\tilde{\psi}\right\rangle=
\lambda \,x_{0}\,,\\
&\Delta x^2=\lambda^2 \Delta x_{0}^2+
\frac{x_{0}^2}{q_{0}^2 k_{0}^2} \Delta p_{0}^2\,,\quad
\Delta p^2=
\frac{1}{\lambda^2} \Delta p_{0}^2\,.
\end{align*}

 $\psi(x,0)$ was assumed to be zero for $x\,>\,0$.
If this function is sufficiently peaked around
$x=x_{0}\,<\,0$ and $p_{0}$, then also $\Delta x$ will be sufficiently small so that
the deformed wave function will also vanish
for $x\,>\,0$, and

\begin{subequations}
\begin{align}
&\int_{-\infty}^{\infty}
 \frac{\kappa}{ \sqrt{2 \pi t \,i}} \frac{1}{\sqrt{2 \pi \hbar }}
\int_{0}^{\infty}  e^{\frac{i (x+y)^2 \kappa ^2}{2 t}}
e^{\frac{i Z y}{\hbar}}
dy\,f(p) dp= 
\\
& \frac{\kappa}{\sqrt{2 \pi t \,i}} 
\int_{-\infty}^{0}  e^{\frac{i (x+y)^2 \kappa ^2}{2 t}}
\tilde{\psi}(y,0) dy \approx 0 \,.
\end{align}
\end{subequations}

Therefore we find
\[
\int_{-\infty}^{\infty}K(x,p,t)f(p)dp
\approx
  \frac{\kappa}{\sqrt{2 \pi t \,i}} 
\int_{-\infty}^{0}  e^{\frac{i (x-y)^2 \kappa ^2}{2 t}}
\tilde{\psi}(y,0) dy\,.
\]
Inserting this result in (\ref{StufegNa}),applying (\ref{Approx}) for the convolution integral and 
proceeding further as for $x\,<\,0$, we get  for $x>0$

 \begin{equation}
 \label{C2}
 \psi(x,t)\approx
 (1+R(p_{0}))
  \frac{\kappa}{\sqrt{2 \pi t \,i}} 
\int_{-\infty}^{0}  e^{\frac{i (x-y)^2 \kappa ^2}{2 t}}
\tilde{\psi}(y,0) dy\, .
\end{equation}

\begin{figure}[htp]
\centering
\subfigure
[The  relative dimensionless probability density $|\Psi|^2 \cdot (\pi \alpha)^{1/2}$ of the initial wave packet.]
{\label{Bild1} 
\includegraphics[width=0.7\textwidth]{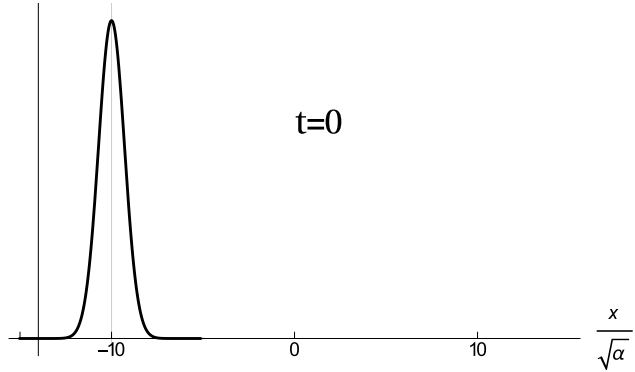}}
\subfigure[The relative dimensionless probability density $|\Psi|^2 \cdot (\pi \alpha)^{1/2}$ at $t_{R}=|x_{0}| m/p_{0}$.]
{\label{Bild2} 
\includegraphics[width=0.7\textwidth]{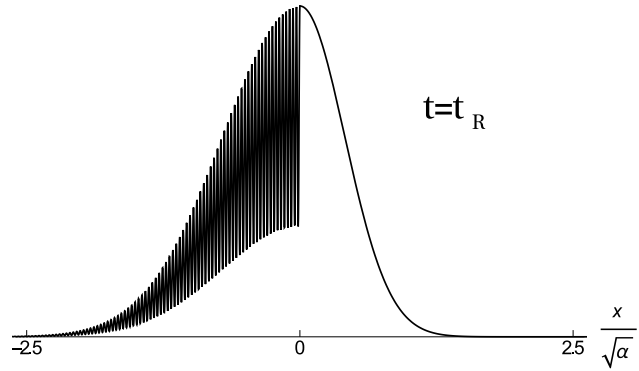}}
\subfigure[The relative dimensionless probability density $|\Psi|^2 \cdot (\pi \alpha)^{1/2}$ at $t=2t_{R}$. The momentum is reduced by a factor
$\sqrt{\frac{k_{0}-1}{k_{0}}}=0.577$.]
{\label{Bild3} 
\includegraphics[width=0.7\textwidth]{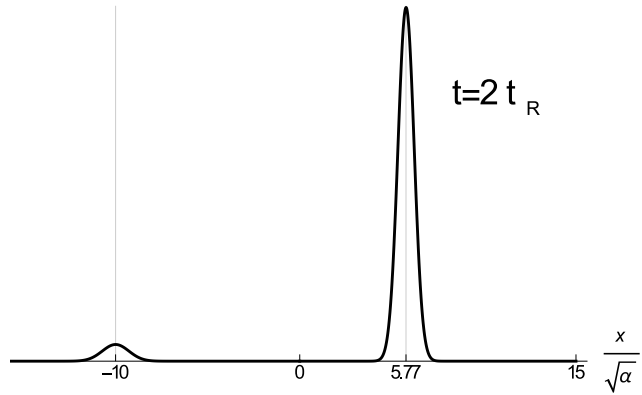}}
\caption{As initial function we choose a Gaussian wave packet with the properties $\Delta x_{0}=\sqrt{\alpha/2}, 
\,\Delta p_{0}=\hbar/\sqrt{2\alpha },\,
x_{0}=-10 \sqrt{\alpha},\,p_{0}=100 \hbar/\sqrt{\alpha}$. The contributions of this 
Gaussian in the region
$x\geq0 $ are so small that we can use (\ref{C1},\ref{C2}) and we can extend the integrals
to the whole real line. We show the wave packet before, during and after the reflection/transmission process
that takes place around the classical reflection time $t_{R}$. We have chosen
$k_{0}=1.5$.}

\label{Climbing}
\end{figure}

This means that the wave function is deformed after it has passed the potential
step. It takes now the form of a free particle wave function that
looked like $\tilde{\psi}(x,0)$ at $t=0$. Moreover it is 
slowed down, and has now the momentum expectation value $\sqrt{p_{0}^2-2 m V}$ instead of $p_{0}$, 
which exactly coincides 
with the reduced momentum
of a corresponding classical particle (see figure \ref{Bild3}).

When the reflection/transmission process is finished
\[
t\,\gg \frac{x_{0} m}{p_{0}}\,,
\]
 only the second term of
(\ref{C1}), describing the reflected wave function is relevant and
the solution consists of a reflected and a transmitted wave packet.

\begin{align}
&\psi(x,t)\approx \psi_{R}(x,t)+\psi_{T}(x,t)= \\
&\quad R(p_{0})
\frac{\kappa}{\sqrt{2 \pi t \,i}} 
\int_{-\infty}^{0}  e^{\frac{i (x+y)^2 \kappa ^2}{2 t}}\psi(y,0) dy+
(1+R(p_{0}))
  \frac{\kappa}{\sqrt{2 \pi t \,i}} 
\int_{-\infty}^{0}  e^{\frac{i (x-y)^2 \kappa ^2}{2 t}}
\tilde{\psi}(y,0) dy
\end{align} 

Here we do not need to restrict $\psi_{R}$
and $\psi_{T}$ to the regions left and right of $x=0$ anymore, since they
will be concentrated in the respective regions anyway.
The reflected and the transmitted part then
fulfill
\[
\left\langle \psi_{R}|\psi_{R}\right \rangle=
(R(p_{0}))^2\,,\quad
\left\langle \psi_{T}|\psi_{T}\right \rangle=
(R(p_{0})+1)^2 \lambda=1-\left\langle \psi_{R}|\psi_{R}\right \rangle
\]
The reflection probability  $\left|R(p_{0}\right|^2$ is then an approximation for
(\ref{Con}) for a sufficiently peaked wave packet.
\subsection{A wave packet reflected by the potential step
and swapping into the forbidden region}
We now consider a wave packet represented by (\ref{Darstellung}),
and concentrated within a region $0\,<\,p\,<p_{m}\,<\sqrt{2 m V}$,
so that we can assume
\begin{equation}
\label{Rand2}
 f(p) \approx 0 \quad 
 \mbox{for} 
 \quad p\,>\,p_{m} \quad \mbox{and}
\quad  p\,<\,0\,,
\end{equation}
and
\begin{equation}
\label{Faktor2}
 \frac{1}{\sqrt{1-\frac{p^2}{2 m V}}} = O(1)\quad \mbox{for} 
 \quad 0 \,\leq\, p\,\leq\,p_{m}\,.
\end{equation}
We  also require (\ref{Approxp1}).
Proceeding as in the case $p_{0}\,>\,\sqrt{2 m V}$, we get the same result
(\ref{C1}) for $x<0$
 \begin{equation}
 \label{S1}
 \psi(x,t)\approx
 \frac{\kappa}{\sqrt{2 \pi t \,i}} 
\int_{-\infty}^{0}  e^{\frac{i (x-y)^2 \kappa ^2}{2 t}}\psi(y,0) dy\, + 
R(p_{0})
\frac{\kappa}{\sqrt{2 \pi t \,i}} 
\int_{-\infty}^{0}  e^{\frac{i (x+y)^2 \kappa ^2}{2 t}}\psi(y,0) dy\,.
\end{equation}

According to (\ref{Rk}),
$R(p_{0})$ is a complex function with absolute value 1 for $k_{0}=\frac{p^2_{0}}{2 m}\,<\,1$. 
For $k_{0}\ll 1$, $R(p_{0})$ is approximately -1
and $\psi(x,t)$ is completely reflected at the barrier.

Applying
\[
\mbox{Erfc}[x]=2-\mbox{Erfc}[-x]
\]
to the second term in (\ref{StufegNb}), and using
(\ref{Error}) with
\[
z=\sqrt{\frac{-i m}{2 h t}} \,,\quad
u=\mp \frac{i m}{\hbar t}x - \frac{i}{\hbar}Z\,,
\]
we get
 \begin{align*}
 & K(x,p,t)=e^{-\frac{i V t}{\hbar}}
 \left\{ \frac{1}{\sqrt{2 \pi \hbar}}
 e^{-\frac{i t Z^2}{2 \hbar m}+\frac{i x Z}{\hbar}}+
 \right. \\
 & \left.
 \frac{\kappa}{ \sqrt{2 \pi t \,i}} \frac{1}{\sqrt{2 \pi \hbar }}
\int_{0}^{\infty}  e^{\frac{i (x+y)^2 \kappa ^2}{2 t}}
e^{\frac{i Z y}{\hbar}}
dy\, - \,
\frac{\kappa}{ \sqrt{2 \pi t \,i}} \frac{1}{\sqrt{2 \pi \hbar }}
\int_{0}^{\infty}  e^{\frac{i (x-y)^2 \kappa ^2}{2 t}}
e^{\frac{i Z y}{\hbar}}dy
\right\}\, .
\end{align*}
With the help of the mean value theorem and (\ref{Approxp1}),(\ref{Faktor2})  we determine
\begin{align*}
&e^{\frac{i Z y}{\hbar}}f(p)=
e^{-\frac{ \sqrt{2 m V-p^2} y}{\hbar}}f(p)=
e^{-\frac{ \sqrt{2 m V-p_{0}^2} y}{\hbar}}f(p)-
e^{-\frac{ \sqrt{2 m V-p_{1}^2} y}{\hbar}}
\frac{p_{1} y}{\hbar \sqrt{2 m V-p_{1}^2}}(p-p_{0})f(p) \\
& \qquad
\approx e^{\frac{- \sqrt{2 m V-p_{0}^2} y}{\hbar}}f(p)\,
\mbox{with}\quad p_{1}\,\epsilon\,(p_{0},p).
\end{align*}

Since according to (\ref{Rand2})
\[
\int_{0}^{\sqrt{2 m V}}f(p) dp\approx
\int_{-\infty}^{\infty}f(p) dp=\psi(0,0)=0\,,
\]
we conclude 
\[
\int_{-\infty}^{\infty}K(x,p,t)f(p)dp
\approx
e^{-\frac{ \sqrt{2 m V-p_{0}^2} x}{\hbar}}
\frac{1}{\sqrt{2 \pi \hbar}} 
\int_{-\infty}^{\infty}  e^{\frac{- i p^2 t}{2 m \hbar}}
f(p) dp\,.
\]
Proceeding as for (\ref{C2}), we finally determine the solution
for $x\,>\,0$

\begin{equation}
\label{S2}
\psi(x,t)\approx (1+R(p_{0}))
e^{-\frac{ \sqrt{2 m V-p_{0}^2} x}{\hbar}}
\frac{1}{\sqrt{2 \pi \hbar}} 
\int_{-\infty}^{\infty}  e^{\frac{- i p^2 t}{2 m \hbar}}
f(p) dp\,.
\end{equation}

We can again conclude  that after the reflection process, when
\[
t\,\gg \frac{x_{0} m}{p_{0}}\,,
\]
the wavefunction only consists of a reflected and a transmitted part.

\begin{align}
&\psi(x,t)\approx \psi_{R}(x,t)+\psi_{T}(x,t)= \\
&\quad R(p_{0})
\frac{\kappa}{\sqrt{2 \pi t \,i}} 
\int_{-\infty}^{0}  e^{\frac{i (x+y)^2 \kappa ^2}{2 t}}\psi(y,0) dy+
(1+R(p_{0}))e^{-\frac{ \sqrt{2 m V-p_{0}^2} x}{\hbar}}
\frac{1}{\sqrt{2 \pi \hbar}} 
\int_{-\infty}^{\infty}  e^{\frac{- i p^2 t}{\hbar}}
f(p) dp\,.
\end{align} 

\begin{figure}[htp]
\centering
\subfigure[The relative dimensionless probability density $|\Psi|^2 \cdot (\pi \alpha)^{1/2}$ of the initial wave packet.]
{\label{Bild4} 
\includegraphics[width=0.7\textwidth]{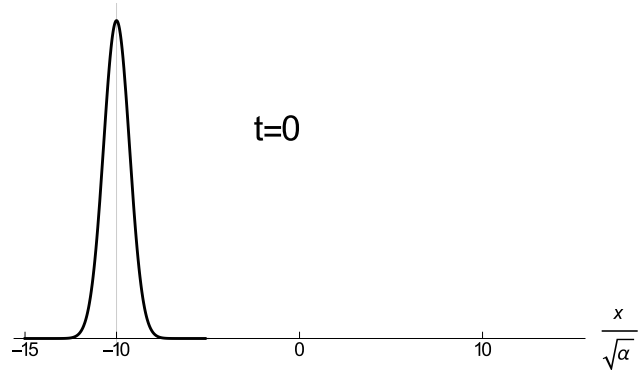}}
\subfigure[The relativeve dimensionless density $|\Psi|^2 \cdot (\pi \alpha)^{1/2}$ at $t_{R}=|x_{0}| m/p_{0}$. Note that the exponential decay
in the classically forbidden region is determined by the factor
$e^{-\sqrt{1/k_{0}-1} p_{0} x/\hbar}$]
{\label{Bild5} 
\includegraphics[width=0.7\textwidth]{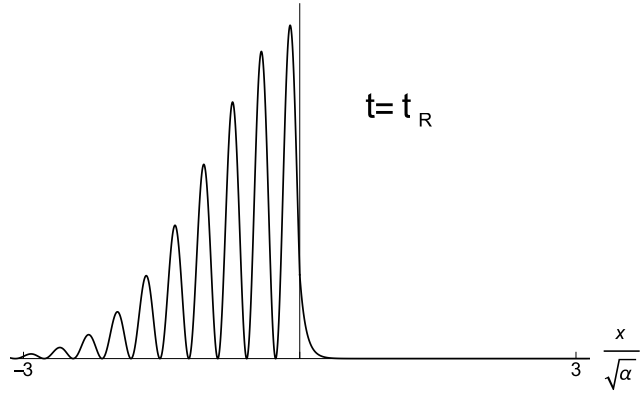}}
\subfigure[The relative dimensionless probability density $|\Psi|^2 \cdot (\pi \alpha)^{1/2}$ at $t=2t_{R}$. The whole wave function  is reflected.
There is no part of the probability density left in the classically forbidden region.]
{\label{Bild6} 
\includegraphics[width=0.7\textwidth]{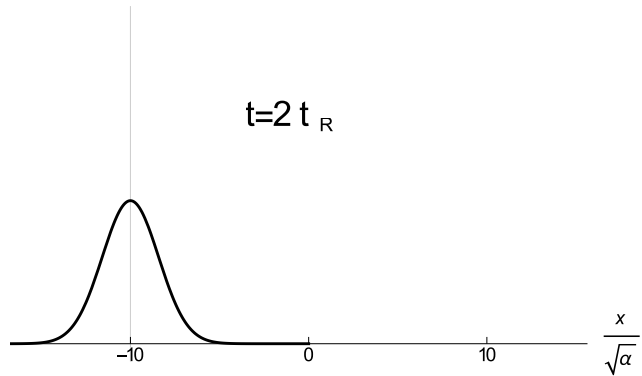}}
\caption{As initial function we choose a Gaussian wave packet with the properties $\Delta x_{0}=\sqrt{\alpha/2}, 
\,\Delta p_{0}=\hbar/\sqrt{\alpha 2},\,
x_{0}=-10 \sqrt{\alpha},\,p_{0}=10 \hbar/\sqrt{\alpha}$.
The contributions of this function for
$x\geq0 $ can again be neglected. 
We see that the wave packet swaps into the region $x>0$ around the classical 
reflection time, but leaves it entirely when the reflection process is over. We have chosen
$k_{0}=1/4$.}
\label{Tunnel}
\end{figure}

But since $|R(p_{0})|=1$, we find
\[
\left\langle\psi_{R}|\psi_{R}\right\rangle=
\int_{-\infty}^{0} \psi^{*}(x,t)\psi(x,t)dx\approx
\int_{-\infty}^{\infty} \psi^{*}(x,t)\psi(x,t)dx=1\quad
\,,
\]
which means that when the reflected wave packet has left the
neighborhood of the step the rest of the wave function also leaves the 
classically forbidden region (see figure \ref{Bild6})
\[
\psi(x,t)\approx \psi_{R}(x,t)\quad
\mbox{for}
\quad t\,\gg \frac{x_{0} m}{p_{0}}\,.
\]

\section{The asymmetric square well}
\label{ASW}

We will consider a potential that is a combination of the infinite square well and the
potential step, which can be described as a box with exit.

\[
\begin{array}{lll}
V(x)=\infty &\qquad \mbox{for} \qquad & x\,\leq\,-d\\
V(x)=0 &\qquad \mbox{for} \qquad &-d\,< x\,<\,0
\\  V(x)=V &\qquad  \mbox{for} \qquad & x\,\geq\,0 \quad.
 \end{array}
\]

The dynamics of  the wave packet $ \psi(x,0) $, is then governed by

\begin{align}
\label{Asa}
&-\frac{\hbar^2}{2 m} \frac{\partial^2 \psi(x,t)}{\partial x^2}
 = i  \hbar \,   \frac{\partial \psi(x,t)}{\partial t} \qquad &
 \mbox{for} \qquad -d\,\leq x\,<\,0 \\
& -\frac{\hbar^2}{2 m} \frac{\partial^2 \psi(x,t)}{\partial x^2}+V \psi(x,t)
 = i  \hbar \,   \frac{\partial \psi(x,t)}{\partial t} \qquad &
 \mbox{for} \qquad  x\,\geq\,0\,,
 \end{align}
 
 where $\psi(x,t)$ is supposed to be continuously differentiable and square integrable
 and to fulfill
 the boundary condition
 \begin{equation}
 \label{boundaryAs}
  \psi(-d,t)=0\,.
 \end{equation}

We will further assume that the initial wave packets
  are located within the box,
 
 \begin{equation}
 \label{Kompakt}
  \psi(x,0)=0 \qquad \mbox{for}\qquad x \geq 0\,.
 \end{equation}

We find for the Laplace transformed wave packet $\varphi(x,s)$

\begin{subequations}
\label{LoesungAs}
\begin{align}
\varphi(x,s)&=\sqrt{\frac{m}{2 i s \hbar}} \left \{u_{1}(x,s)\int_{-d}^{x}u_{2}(y,s) 
\psi(y,0) dy \,  - \,\right.&
\left.
 u_{2}(x,s) 
\int_{-d}^{x}u_{1}(y,s) \psi(y,0) dy 
\right\} \nonumber \\
& &\nonumber \\
&\quad +\, \alpha(s) u_{1}(x,s)  \,+\, \beta(s) u_{2}(x,s) 
 &\mbox{for}  \quad x\,<\,0 \qquad\\
& &\nonumber \\
\varphi(x,s)&= \gamma(s) u_{3}(x,s)  \,+\, \delta(s) u_{4}(x,s)
 &\mbox{for}  \quad x\,>\,0 \qquad ,
\end{align}
\end{subequations}

where the functions $u_{1},u_{2},u_{3},u_{4}$ are defined as in the previous section (\ref{u1234}).

Since  $\varphi(x,s)$ must vanish for $x\rightarrow \infty$, we find
\[
 \delta(s)=0\,.
\]
 The boundary condition (\ref{boundaryAs}) implies 
 \[
  \alpha(s)=-\beta(s) e^{(i-1) 2 d \kappa \sqrt{s}}\,.
 \]
 Since $\varphi(x,s)$ should be continuously differentiable at $x=0$, we get 
 
 \begin{align}
 & \gamma(s)=\frac{\kappa}{\sqrt{2 \,s\, i}}\cdot
  \frac{2 \,I_{2}-2 \,I_{1}e^{(i-1)2 d \kappa \sqrt{s}}}
  {1+\sqrt{1-\frac{V}{s i \hbar}}+e^{(i-1)2 d \kappa \sqrt{s}}
  \left(1-\sqrt{1-\frac{V}{s i \hbar}}\right)} \\
  & \beta(s)=\frac{\kappa}{\sqrt{2 \,s\, i}}\cdot\frac{1}{1-e^{(i-1)2 d \kappa \sqrt{s}}}
  \left\{
  I_{1}-I_{2}+ \frac{2 \,I_{2}-2 \,I_{1}e^{(i-1)2 d \kappa \sqrt{s}}}
  {1+\sqrt{1-\frac{V}{s i \hbar}}+e^{(i-1)2 d \kappa \sqrt{s}}
  \left(1-\sqrt{1-\frac{V}{s i \hbar}}\right)}
  \right\} \,,
  \end{align}
  where we have introduced
  \[
   I_{1}=\int_{-d}^{0}e^{(i-1) y \kappa \sqrt{s}}\psi(y,0)dy\,,\quad
   I_{2}=\int_{-d}^{0}e^{-(i-1) y \kappa \sqrt{s}}\psi(y,0)dy\,.
   \]

Inserting these results in \ref{LoesungAs}, using the abbreviations (\ref{LKoeffizient},\ref{Kappa}) and
applying 
\[
\frac{1-\sqrt{1-\frac{V}{s i h}}}{1+\sqrt{1-\frac{V}{s i h}}}=
\rho(s)
\]

as well as the series expansion
\begin{equation}
\frac{1}{(1+e^{ (i-1)\kappa 2 d\sqrt{s}} \rho(s))}=
\sum_{k=0}^{\infty} (-1)^k \rho(s)^{k} \,e^{ (i-1)2 d \kappa \sqrt{s}\, k}\,,
\end{equation}
we find for the Laplace transformed wave packet  (\ref{LoesungAs}),

\begin{subequations}
\label{LAs}
 \begin{align}
  \label{LAsa}
 & \varphi(x,s)=\left\{
  \sum_{k=0}^{\infty}\frac{\kappa }{\sqrt{2 s i }}
  (\rho(s)+1) (-1)^k\rho(s)^{k}
  \int_{-d}^{0}e^{(i-1)(2 d k-y)\kappa \sqrt{s}}\psi(y,0)dy \right. \nonumber\\
  & \quad
  \left. - \sum_{k=0}^{\infty}\frac{\kappa }{\sqrt{2 s i }}
  (\rho(s)+1) (-1^k)\rho(s)^{k}
  \int_{-d}^{0}e^{(i-1)(2 d (k+1)+y)\kappa \sqrt{s}}\psi(y,0)dy \right\}
  e^{i\sqrt{\frac{2 m s i}{\hbar}-\frac{2 m V}{\hbar^2}} x} \nonumber \\
  &\qquad \mbox{for} \quad x\,>\,0\,.
 \end{align}
 
 Applying also
 \begin{align*}
 & \frac{1}{1+e^{(i-1)2 d \kappa \sqrt{s}}\rho(s)}\cdot \frac{1}{1-e^{(i-1)2 d \kappa \sqrt{s}}}
  =\frac{e^{-(i-1)2 d \kappa \sqrt{s}}}{1+\rho(s)}
  \left\{
  \frac{1}{1-e^{(i-1)2 d \kappa \sqrt{s}}}- \frac{1}{1+e^{(i-1)2 d \kappa \sqrt{s}}\rho(s)}
  \right\}
  \\
  &\quad=\frac{1}{1+\rho(s)}\sum_{k=0}^{\infty}
  (1-(-1)^k\rho(s)^{k})e^{(i-1)2 d (k-1)\kappa \sqrt{s}}\,,
 \end{align*}

 we conclude
 \begin{align}
   \label{LAsb}
  & \varphi(x,s)=
  -\sum_{k=0}^{\infty}\rho(s)^k (-1)^k
  \frac{\kappa }{\sqrt{2 s i }}
  \int_{-d}^{0}
  e^{(i-1)(2 d (k+1)+x+y)\kappa \sqrt{s}}\psi(y,0)dy \nonumber \\
  & \quad
  +\sum_{k=0}^{\infty}\rho(s)^{k+1} (-1)^{k+1}
 \frac{\kappa }{\sqrt{2 s i }}
  \int_{-d}^{0}
  e^{(i-1)(2 d (k+1)+x-y)\kappa \sqrt{s}}\psi(y,0)dy \nonumber \\
   & \quad + \frac{\kappa }{\sqrt{2 s i }}\int_{-d}^{0}
  e^{(i-1)|x-y|\kappa \sqrt{s}}\psi(y,0)dy \nonumber \\
   & \quad
  +\sum_{k=0}^{\infty}\rho(s)^{k+1} (-1)^{k+1}
 \frac{\kappa }{\sqrt{2 s i }}
  \int_{-d}^{0}
  e^{(i-1)(2 d (k+1)-x+y)\kappa \sqrt{s}}\psi(y,0)dy \nonumber \\
   & \quad
  -\sum_{k=0}^{\infty}\rho(s)^{k+2} (-1)^{k+2}
  \frac{\kappa }{\sqrt{2 s i }}
  \int_{-d}^{0}
  e^{(i-1)(2 d (k+1)-x-y)\kappa \sqrt{s}}\psi(y,0)dy \nonumber \\
  & \quad + \frac{\kappa }{\sqrt{2 s i }}
  \rho(s)
  \int_{-d}^{0}
  e^{(i-1)|x+y|\kappa \sqrt{s}}\psi(y,0)dy \nonumber \\
  &\qquad \mbox{for} \quad x\,<\,0\,.
 \end{align}
 \end{subequations}
In order to determine the inverse Laplace transform of (\ref{LAsa}),
we will proceed similar as in section (\ref{GN}).
We will use the shifted momentum representation

\begin{equation}
\label{SMREp}
 f(K,p)=\frac{1}{\sqrt{2 \pi \hbar}}\int_{-\infty}^{\infty} \psi(x+K,0) e^{\frac{-i p x}{\hbar}} dx\,=\,
 e^{\frac{i p K}{\hbar}}f(p)
\end{equation}

and we introduce
\begin{equation}
\label{PKoeffizient}
L(k,t)=\mathcal{L}^{-1}\left\{(\rho(s)+1)\rho(s)^k\right\}\,,
\end{equation}

which can be expressed as a sequence of convolutions of
(\ref{Koeffizient}).

\begin{subequations}
This yields for the wave function in the region $x\,>\,0$
\begin{align}
\label{AsgN}
&\psi(x,t)=\int_{-\infty}^{\infty} K\left(x, p, t\right)f(0,p)\,dp+
 \int_{0}^{\infty} \int_{-\infty}^{\infty} K\left(x, p ,t-\tau\right)f(0,p)\,dp\,r(\tau)\,d\tau\, +\nonumber \\
 &\quad
 \sum_{k=1}^{\infty} (-1)^k
 \int_{0}^{t} \int_{-\infty}^{\infty}
 K\left(x, p, t-\tau\right)f(2 d k,p)\,dp\, L(k,\tau) d\tau 
 \nonumber \\
& \quad
- \int_{-\infty}^{\infty} K\left(x, p, t\right)f(-2d,p)\,dp-
 \int_{0}^{\infty} \int_{-\infty}^{\infty} K\left(x, p ,t-\tau\right)f(-2d,p)\,dp\,r(\tau)\,d\tau\,
 \nonumber\\
 &\quad
 -(-1)^k\sum_{k=1}^{\infty}
 \int_{0}^{t} \int_{-\infty}^{\infty}
 K\left(x, p, t-\tau\right)f(-2 d k,p)\,dp\, L(k,\tau) d\tau\,.
\end{align}

Note that the first two terms are identical with the solution of the potential step (\ref{StufegN}). They describe the behavior
of a wave packet that starts with positive momentum and undergoes
its first transmission process. The following sum describes the transmitted
parts of this wave packet after the kth reflection at the left wall $x=-d$.
The remaining terms  are relevant for the description of an initial wave packet with negative momentum.
\end{subequations}

The inverse Laplace transform of (\ref{LAsb}) yields

\begin{align}
\label{AskN}
&\psi(x,t)=
-\sum_{k=0}^{\infty}
\int_{0}^{t}\int_{-d}^{0}
\frac{\kappa}{\sqrt{2 \pi i (t-\tau)}}
e^{\frac{i \kappa^2 (2d(k+1)+x+y)^2}{2 (t-\tau)}}
\psi(y,0)dy \,(-1)^k M(k,\tau) d\tau \nonumber \\
& \qquad +\sum_{k=0}^{\infty}
\int_{0}^{t}\int_{-d}^{0}
\frac{\kappa}{\sqrt{2 \pi i (t-\tau)}}
e^{\frac{i \kappa^2 (2d(k+1)+x-y)^2}{2 (t-\tau)}}
\psi(y,0)dy \,(-1)^{k+1} M(k+1,\tau) d\tau \nonumber \\
& \qquad +\sum_{k=0}^{\infty}
\int_{0}^{t}\int_{-d}^{0}
\frac{\kappa}{\sqrt{2 \pi i (t-\tau)}}
e^{\frac{i \kappa^2 (2d(k+1)-x+y)^2}{2 (t-\tau)}}
\psi(y,0)dy \,(-1)^{k+1} M(k+1,\tau) d\tau \nonumber \\
& \qquad -\sum_{k=0}^{\infty}\sum_{l=0}^{\infty}
\int_{0}^{t}\int_{-d}^{0}
\frac{\kappa}{\sqrt{2 \pi i (t-\tau)}}
e^{\frac{i \kappa^2 (2d(k+1)-x-y)^2}{2 (t-\tau)}}
\psi(y,0)dy \,(-1)^{k+2} M(k+2,\tau) d\tau \nonumber \\
&\qquad +\int_{-d}^{0}
\frac{\kappa}{\sqrt{2 \pi i t}}
e^{\frac{i \kappa^2 (x-y)^2}{2 t}}
\psi(y,0)dy \,+\,
\int_{0}^{t}\int_{-d}^{0}
\frac{\kappa}{\sqrt{2 \pi i (t-\tau)}}
e^{\frac{i \kappa^2 (x+y)^2}{2 (t-\tau)}}
\psi(y,0)dy \,r(\tau) d\tau \nonumber \\
&\qquad \mbox{for} \quad x\,<\,0 \,,
\end{align}

where we have introduced

\begin{equation}
\label{PKoeffizient2}
M(k,t)\equiv\mathcal{L}^{-1} \left( \rho(s)^k \right)\,.
\end{equation}

The last two terms of (\ref{AskN})coincide with the solution of the potential step (\ref{StufekN}). 
If we compare (\ref{AskN}) with
the solution of the infinite square well (\ref{DynKb}), we find
that (\ref{AskN}) consists of the terms of (\ref{DynKb}) and their convolutions
with $M(k,t)$.
If $ t \gg 2\hbar/V $ and as long as only a limited number of terms are involved, the 
convolution integral can be extended to infinity as for the approximate solutions of the potential step
(see Appendix \ref{Aexit}).
 We find for a sufficiently peaked wave packet with momentum
expectation value $p_{0}$

\begin{align}
\label{Exit}
&\psi(x,t)\approx
-\sum_{k=0}^{L_{1}}(-1)^k R(p_{0})^k
\int_{-d}^{0}
\frac{\kappa}{\sqrt{2 \pi i t}}
e^{\frac{i \kappa^2 (2d(k+1)+x+y)^2}{2 t}}
\psi(y,0)dy 
\nonumber \\
& \qquad +\sum_{k=0}^{L_{2}}
 (-1)^{k+1} R(p_{0})^{k+1}\int_{-d}^{0}
\frac{\kappa}{\sqrt{2 \pi i t}}
e^{\frac{i \kappa^2 (2d(k+1)-x+y)^2}{2 t}}
\psi(y,0)dy 
\nonumber \\
& \qquad +\sum_{k=0}^{L_{3}}(-1)^{k+1} R(p_{0})^{k+1}
\int_{-d}^{0} 
\frac{\kappa}{\sqrt{2 \pi i t}}
e^{\frac{i \kappa^2 (2d(k+1)+x-y)^2}{2 t}}
\psi(y,0)dy 
\nonumber \\
& \qquad -\sum_{k=0}^{L_{4}} (-1)^{k+2} R(p_{0})^{k+2}
\int_{-d}^{0}
\frac{\kappa}{\sqrt{2 \pi i t}}
e^{\frac{i \kappa^2 (2d(k+1)-x-y)^2}{2 t}}
\psi(y,0)dy 
\nonumber \\
&\qquad +\frac{\kappa}{\sqrt{2 \pi t \,i}} 
\int_{-\infty}^{0}  e^{\frac{i (x-y)^2 \kappa ^2}{2 t}}\psi(y,0) dy\, + 
R(p_{0})
\frac{\kappa}{\sqrt{2 \pi t \,i}} 
\int_{-\infty}^{0}  e^{\frac{i (x+y)^2 \kappa ^2}{2 t}}\psi(y,0) dy
\nonumber \\
&\qquad \mbox{for} \quad x\,<\,0 \,.
\end{align}

We find that the approximate solution of the asymmetric square
well looks like the solution of the infinite well (\ref{DynKb})
supplemented by the appropriate reflection coefficients that
make allowance for the fact that at each time the wave packet
is reflected at $x=0$, parts of the wave packet leave the box.
If the wave packet is located between the walls (not too close to any of them) it is  described
by only term of (\ref{AskN}). The probability of finding a corresponding particle within the wall is
then given by $\left|R(p_{0})\right|^{2m}$, where $m$ denotes the already undergone number of reflections
at the right wall. The transmission probability is then given by $1-\left|R(p_{0})\right|^{2m}$.

\section{Conclusions}

We applied the  method of Laplace transform to reproduce the mirror solution for
the infinite square well in a straightforward way and to derive exact and intuitive solutions
for the potential step and the asymmetric square well.

We could show that a wave packet with energy higher  than
the potential step will partly be reflected and partly transmitted
and thereby slowed down (\ref{Bild2}), as it would be expected
from a classical particle that has passed the step.
We found that the wave function with energy lower than the potential step
will swap into the classically forbidden region only during the reflection process. 
Moreover we derived an inequality for the reflection probability in terms of the 
reflection coefficient that is saturated for certain initial wave packets (\ref{Finding},\ref{Con}).

In the case of the asymmetric well the solution inside the well (\ref{AskN}) looks like a 
generalization of the solution of the infinite well (\ref{DynKb}). Each term of (\ref{AskN}) 
describes the wave packet
after a certain number of reflections and the convolution integrals can be 
approximated by powers of the reflection coefficient if the
wave packet is sufficiently narrow (\ref{Exit}) as long as the system still shows 
semiclassical behaviour. For this case we also determined the probability of finding a particle 
inside the box at
a given time which yields the corresponding transmission probability.
Note that this example illustrates
the advantages of the method very well since from the point
of view of the characterization in terms of eigenstates the box with exit differs fundamentally from 
the infinite square well since there are bound and unbound eigenstates.

From the technical point of view we needed for the box with exit only the  inverse Laplace 
 transformation
 of functions that were already used for the infinite square well and the potential step
 (\ref{Wurzel1}, \ref{Koeffizient}). We expect that the solutions of all piecewise 
 constant potentials can be characterized by convolutions of these inverse Laplace transformed
 functions since the asymmetric square well is a generic case with no special symmetry and a finite as 
 well as an infinite wall. The explicite form of the solutions make the method especially 
 useful for the studying
 of the tunneling time. It would also be interesting to obtain time dependent solutions for other sorts
 of exactly solvable potentials 
 as the Morse oscillator or the Hydrogen atom.
A generalization of the method to more spatial dimensions can be achieved by using the Duhamel principle.

\appendix
\section{The deformed wave function}
\label{Deformed}
The substitution $q=\sqrt{p^2-2 m V}$, 
$q_{0}=\sqrt{p_{0}^2-2 m V}$ yields
\begin{align*}
&\tilde{\psi}(y,0)=
\frac{1}{\sqrt{2 \pi  \hbar}}
\int_{\sqrt{2 m V}}^{\infty}
e^{\frac{i \sqrt{p^2-2 m V} y}{\hbar}}f(p)dp= \\
& \qquad\frac{1}{\sqrt{2 \pi  \hbar}}
\int_{0}^{\infty}
e^{\frac{i q y}{\hbar}}f(\sqrt{q^2+2 m V})\frac{q}{\sqrt{q^2+2 m V}}dq= \\
& \qquad
\frac{1}{\sqrt{2 \pi  \hbar}}
\int_{0}^{\infty}
e^{\frac{i q y}{\hbar}}
e^{-\frac{i \sqrt{q^2+2 m V}x_{0}}{\hbar}}
F\left(\sqrt{q^2+2 m V}-\sqrt{q_{0}^2+2 m V}\right)
\frac{q}{\sqrt{q^2+2 m V}}dq\,.
\end{align*} 

Applying(\ref{Approxp2})and expanding also $q/\sqrt{q^2_{0}+2 m V}$ around $q_{0}$ and using
(\ref{Approxp1}),
we find
\[
\tilde{\psi}(y,0)\approx
\frac{1}{\sqrt{2 \pi h}}\int_{0}^{\infty}
e^{\frac{i q y}{\hbar}}
\lambda e^{-\frac{i q_{0 }x_{0}}{\lambda \hbar}
-\frac{i x_{0}\lambda}{ \hbar}(q-q_{0 })
-\frac{i x_{0}\lambda}{2 \hbar q_{0}k_{0}}(q-q_{0 })^2}
F\left(\lambda(q-q_{0})\right)
dq\,.
\]

Because of (\ref{Rand}) we can extend the integral
\[
\tilde{\psi}(y,0)\approx
\frac{1}{\sqrt{2 \pi h}}\int_{-\infty}^{\infty}
e^{\frac{i q y}{\hbar}}
\lambda e^{-\frac{i q_{0 }x_{0}}{\lambda \hbar}
-\frac{i x_{0}\lambda}{ \hbar}(q-q_{0 })
-\frac{i x_{0}\lambda}{2 \hbar q_{0}k_{0}}(q-q_{0 })^2}
F\left(\lambda(q-q_{0})\right)
dq\,.
\]
Note that in (\ref{Approxp2})the argument of $F$ is approximated 
by a Taylor expansion around $q_{0}$ up to first order, whereas the exponent is expanded to second order 
due to a factor $1/\hbar$
\section{Approximation for the asymmetric square well}
\label{Aexit}

The solution (\ref{AskN}) consists of terms of the form

\begin{align}
& U(k,t)=\int_{0}^{t}\frac{\kappa}{\sqrt{2 \pi (t-\tau) \,i}} 
\int_{-d}^{0} e^{\frac{i (Q\pm y)^2 \kappa ^2}{2 (t-\tau)}} \psi(y,0) dy\,M(k,\tau)d\tau
= \\
&\int_{0}^{t}\frac{1}{\sqrt{2 \pi \hbar}} \int_{-\infty}^{\infty} e^{\frac{\mp i p Q}{\hbar}}
e^{-\frac{i p^2 (t-\tau)}{2m \hbar}}f(p)  dp M(k,\tau) d\tau\quad
\mbox{with} \quad Q=2dk\pm x
\end{align}
where, according to (\cite{Erdely}) 

\begin{equation}
M(k,t)=\mathcal{L}^{-1} \left( \rho(s)^k \right)\,
=\frac{k}{i^k t}J_{k}\left[\frac{V t}{2 h}\right]e^{-\frac{i V t}{2 h}}.
\end{equation}
If we approximate the $\tau$-integral by an infinite Laplace integral as for (\ref{StufekNM}), 
we find

\begin{align}
& U(k,t) \approx
 \int_{0}^{\infty}\frac{1}{\sqrt{2 \pi \hbar}} \int_{-\infty}^{\infty} 
 e^{\frac{\mp i p Q}{\hbar}}
e^{-\frac{i p^2 (t-\tau)}{2m \hbar}}f(p)  dp M(k,\tau) d\tau= \nonumber \\
& \quad \frac{1}{\sqrt{2 \pi \hbar}} \int_{-\infty}^{\infty} e^{\frac{\mp i p Q}{\hbar}}
e^{-\frac{i p^2 t}{2m \hbar}}f(p) \left(\rho\left[\frac{-i p^2}{2 m h}\right]\right)^k dp=
\nonumber \\
&\quad \frac{1}{\sqrt{2 \pi \hbar}} \int_{-\infty}^{\infty} e^{\frac{\mp i p Q}{\hbar}}
\quad e^{-\frac{i p^2 t}{2m \hbar}}f(p) R(p)^k dp
\label{U}
\end{align}

For the neglected part of the integral 

\begin{align}
\label{u}
u(k,t)=
\int_{t}^{\infty} e^{\frac{\mp i p Q}{\hbar}}
e^{-\frac{i p^2 (t-\tau)}{2m \hbar}} e^{-\frac{i V \tau}{2 h}}\frac{k}{i^k \, \tau}
J_{k}\left[\frac{V \tau}{2 h}\right] d\tau \,,
\end{align}

we get the following estimate
\begin{subequations}
\begin{align}
&\left|
\int_{t}^{\infty}e^{\frac{i p^2 \tau}{2m \hbar}-\frac{i V \tau}{2 h}}\frac{k}{\tau}
J_{k}\left[\frac{V t}{2 h}\right] d\tau
\right|\,\leq
\int_{\frac{t V}{2 \hbar}}^{\infty}
\left|\frac{k\,J_{k}(y)}{y}\right| dy\,\leq \\
&\left(\int_{\frac{t V}{2 \hbar}}^{\infty}\frac{k}{y^{1+2\epsilon}} dy \right)^{\frac{1}{2}}
\cdot
\left(\int_{0}^{\infty}
\frac{\left(J_{k}(y)\right)^2}{y^{1-2\epsilon}}
\right)^{\frac{1}{2}}\,=\,\\
& k \frac{1}{\sqrt{2 \epsilon}} \left(\frac{2 h}{V t}\right)^{\epsilon}
\cdot
\left(
2^{2 \epsilon}
\frac{\Gamma [1-2 \epsilon]\Gamma[k+\epsilon]}
{2 (\Gamma[1-\epsilon])^2 \Gamma[1+k-\epsilon]}
\right)^{\frac{1}{2}} \,\leq \\
\label{est}
& k \frac{1}{\sqrt{2 \epsilon}}\left(\frac{2 h}{V t}\right)^{\epsilon}
\cdot
\left(
2^{2 \epsilon}
\frac{\Gamma [1-2 \epsilon]}
{2 (\Gamma[1-\epsilon])^2 }
\right)^{\frac{1}{2}}
\quad \mbox{with} \quad 0\,<\,\epsilon\,<\,\frac{1}{2}\,,
\end{align}
\end{subequations}

where we have applied the Schwarz inequality and the integral formula
(\cite{Magnus})
\begin{equation}
\label{Bessel2}
\int_{0}^{\infty}
\frac{\left(J_{k}(y)\right)^2}{y^{1-2\epsilon}}=
2^{2 \epsilon}
\frac{\Gamma [1-2 \epsilon]\Gamma[k+\epsilon]}
{2 (\Gamma[1-\epsilon])^2 \Gamma[1+k-\epsilon]}\,.
\end{equation}

Therefor (\ref{U}) will be a good approximation for $U(k,t)$ if
$t\gg \frac{ k^ {\frac{1}{\epsilon}}\hbar}{V }$.
Moreover, if the wave packet is sufficiently peaked around the 
momentum expectation value $p_{0}$, so that $R(p_{0})^k$ can be put before the integral, we 
can write

\begin{align}
& U(k,t) \approx
 R(p_{0})^k 
\frac{1}{\sqrt{2 \pi \hbar}} 
\int_{-\infty}^{\infty} e^{\frac{\mp i p Q}{\hbar}}
\quad e^{-\frac{i p^2 t}{2m \hbar}}f(p)  dp= \\
& \quad R(p_{0})^k 
\frac{\kappa}{\sqrt{2 \pi t \,i}} 
\int_{-d}^{0} e^{\frac{i (Q\pm y)^2 \kappa ^2}{2 t}} \psi(y,0) dy \,.
\end{align}

However, the estimation (\ref{est}) gets worse for increasing $k$ and moreover the solution (\ref{AskN}) 
consists of infinite sums of terms of the form $U(k,t)$  Therefor we can use the  approximation only if 
the number of relevant terms is limited.

We will now show how the approximation works for an initial wave function of the form
\[
 \psi(y,0)=e^{\frac{i p_{0} y}{\hbar}}G(y)\,,
\]
where $G(y)$ is a two times differentiable real function that is zero outside $[0,d]$ and $p_{0}=m v_{0}$ 
denotes the momentum expectation value of $\psi(y,0)$.
Without loss of generality we will only discuss the sum

\begin{subequations}
\begin{align}
&S_{1}=\sum_{k=1}^{\infty}U_{1}(k,t)\,,\mbox{with} \label{Su1}\\
&U_{1}(k,t)=\int_{0}^{t}\frac{\kappa}{\sqrt{2 \pi (t-\tau) \,i}} 
\int_{-d}^{0} e^{\frac{i (2 d k+x- y)^2 \kappa ^2}{2 (t-\tau)}} \psi(y,0) dy\,M(k,\tau)d\tau\,.
\label{U1}
\end{align}
\end{subequations}

Considering only terms with 

\[
 X=\kappa \frac{2 d k-x+y-v_{0}(t-\tau)}{2 \sqrt{t-\tau}}\,>\,0
\]

and applying the estimate \cite{Abramowitz}

\[
 \left|
 -\mbox{Erfc}[(1-i)X] X+\frac{e^{2 i X^2}}{(1-i)\sqrt{\pi}}
 \right|\,\leq\,
 \frac{1}{4 \sqrt{2 \pi} X^2}
\]

we find after a twofold partial integration for the spatial integral

\begin{align}
& \left| \int_{-d}^{0} \frac{\kappa}{\sqrt{2 \pi (t-\tau) \,i}} 
 e^{\frac{i (2 d k+x- y)^2 \kappa ^2}{2 (t-\tau)}}
 e^{\frac{i p_{0} y}{\hbar}}G(y) dy\right|\,= \\
& \left| \int_{-d}^{0} 
 \frac{\sqrt{t-\tau)}}{\kappa}
 \left\{ -\mbox{Erfc}[(1-i)X] X+\frac{e^{2 i X^2}}{(1-i)\sqrt{\pi}}\right\}
 G''(y) dy\right|\,\leq\\
&  \left(\frac{\sqrt{t-\tau)}}{\kappa} \right)^3
\frac{1}{\sqrt{2\pi}} \int_{-d}^{0} 
\frac{1}{(2 d k- v_{0}\left(t-\tau \right)-x+y)^2}\left| G''(y)\right|dy\,\leq \\
&  \left(\frac{\sqrt{t-\tau)}}{\kappa} \right)^3
\frac{1}{\sqrt{2\pi}} \frac{1}{(2 d k- v_{0} t-d)^2}\int_{-d}^{0} 
\left| G''(y)\right|dy\,.
\end{align}

Applying again the Schwarz inequality and (\ref{Bessel2}) for $\epsilon=-1/2$, we can
write for the convolution integral

\begin{align}
 &\left|\int_{0}^{\infty}
 \left(\frac{\sqrt{t-\tau}}{\kappa}\right)^3\frac{k}{i^k \tau}
 J_{k}\left(\frac{\tau V}{2 \hbar}\right) d\tau
 \right|\,\leq\\
& \frac{1}{\kappa^3} \left\{ (t-\tau)^3 d\tau\right\}^{\frac{1}{2}}\cdot
 k 
 \left\{\left(\frac{V}{2 \hbar}\right)
 \int_{0}^{\infty}\frac{J_{k}(y)}{y} dy\right\}^{\frac{1}{2}}=\\
 &\frac{t^2 \hbar}{2 }\sqrt{\frac{V}{2 m^3}}\cdot
 \frac{k}{\sqrt{\pi \left(k^2-\frac{1}{4}\right)}}\,\leq \,
t^2 \hbar\sqrt{\frac{V}{6 \pi m^3}}\,.
 \end{align}

Then we find for the sum of all $U_{1}(k,t)$ (\ref{U1}) with $k\,\leq\,l$, where $l$ 
is the smallest positive integer with $l\, 2 d \geq v_{0} t +3 d$,

\begin{align}
 &\left|\sum_{k=l}^{\infty}U(k,t)\right|\,\leq \,
 \sum_{k=l}^{\infty}\left|U(k,t)\right|\,\leq\, \\
& \frac{1}{\sqrt{2 \pi}} \left(\sum_{m=1}^{\infty}\frac{1}{4d^2 m^2}\right)
 t^2 \hbar\sqrt{\frac{V}{6 \pi m^3}}\int_{-d}^{0} 
\left| G''(y)\right|dy\,\leq
\frac{\pi t^2 \hbar}{24 d^2}\sqrt{\frac{V}{18 \pi m^3}}\int_{-d}^{0} 
\left| G''(y)\right|dy\,,
\end{align}

which can be neglected if 
\begin{equation}
\label{cond1}
 \frac{ t^2 \hbar}{d^3}\sqrt{\frac{V}{ m^3}}\,\ll\,1\,.
\end{equation}

So we can write for (\ref{Su1})

\begin{align}
 &S_{1}\approx\sum_{k=1}^{l-1}U_{1}(k,t)=
\sum_{k=1}^{l-1} \frac{1}{\sqrt{2 \pi \hbar}} \int_{-\infty}^{\infty} 
e^{\frac{\mp i p (2 d k+x)}{\hbar}}
\quad e^{-\frac{i p^2 t}{2m \hbar}}f(p) R(p)^k dp\\
&\qquad +\sum_{k=1}^{l-1} u_{1}(k,t)\,,
\end{align}
where $u_{1}(k,t)$ is defined as a special case of $u(k,t)$ (\ref{u}),namely

\begin{equation}
u(k,t)=
\int_{t}^{\infty} e^{\frac{ i p (2 d k+x)}{\hbar}}
e^{-\frac{i p^2 (t-\tau)}{2m \hbar}} e^{-\frac{i V \tau}{2 h}}\frac{k}{i^k \, \tau}
J_{k}\left[\frac{V \tau}{2 h}\right] d\tau\,. 
\end{equation}

Applying (\ref{est}), we find

\begin{align}
&\left|\sum_{k=1}^{l-1} u_{1}(k,t)\right|\,\leq\,
\sum_{k=1}^{l-1}\left| u_{1}(k,t)\right|\,\leq \\
&\frac{1}{\sqrt{2 \epsilon}}\left(\frac{2 h}{V t}\right)^{\epsilon}
\cdot
\left(
2^{2 \epsilon}
\frac{\Gamma [1-2 \epsilon]}
{2 (\Gamma[1-\epsilon])^2 }
\right)^{\frac{1}{2}} \cdot \frac{l (l-1)}{2}\,.
\end{align}

So this sum can be neglected, if
\begin{equation}
\label{cond2}
\left(\frac{2 h}{V t}\right)^{\epsilon}
\cdot \frac{l (l-1)}{2}\,\ll\,1\,,
\end{equation}

and we finally find 
\begin{equation}
S_{1}\approx
\sum_{k=1}^{L_{1}} \frac{1}{\sqrt{2 \pi \hbar}} \int_{-\infty}^{\infty} 
e^{\frac{\mp i p (2 d k+x)}{\hbar}}
\quad e^{-\frac{i p^2 t}{2m \hbar}}f(p) R(p)^k dp\,,
\end{equation}

if both conditions (\ref{cond1},\ref{cond2}) are fulfilled. Here $L_{1}=l-1$. Similarly
estimates for the other three types of sums in (\ref{AskN}) can be found  that will
limit the number of relevant terms to $L_{2},L_{3},L_{4}$.

\section*{Acknowledgments}

I thank Helmut Rumpf for his careful reading of this article and his valuable advices.
I also thank Beatrix Hiesmayr, Helmuth H\"uffel and Helmuth Urbantke for their helpful comments.

\end{document}